\documentclass[conference]{IEEEtran}
\usepackage{cite}
\usepackage{graphics}
\usepackage{epsfig}
\usepackage{subfigure} 
\usepackage{url} 
\usepackage{amsmath} 
\usepackage{array}
\usepackage{amsbsy}
\usepackage{amssymb}

\newcounter{example}[section]

\begin{document}
\title{Selection Relaying at Low Signal to Noise Ratios}
\author{Ketan Rajawat and Adrish Banerjee\\
Department of Electrical Engineering \\
Indian Institute of Technology Kanpur\\
Kanpur-208016, India \\
Email: adrish@iitk.ac.in} 

\maketitle

\begin{abstract}
Performance of cooperative diversity schemes at Low Signal to Noise Ratios (LSNR) was recently studied by Avestimehr et. al. \cite{baf} who emphasized the importance of diversity gain over multiplexing gain at low SNRs. It has also been pointed out that continuous energy transfer to the channel is necessary for achieving the max-flow min-cut bound at LSNR. Motivated by this we propose the use of Selection Decode and Forward (SDF) at LSNR and analyze its performance in terms of the outage probability. We also propose an energy optimization scheme which further brings down the outage probability.
\end{abstract}

\section{Introduction}
Cooperative diversity has attracted considerable interest of researchers recently. The focus has been on design of efficient protocols, especially in the slow fading scenario \cite{laneman,dmtgamal} where spatial diversity offers an interesting method to combat the occasional deep fades in the channel.

At high Signal-to-Noise ratios (HSNR), the performance of a cooperative diversity protocol is best measured by the diversity-multiplexing trade off it achieves \cite{dmttse, dmtgamal}. At low SNRs however, as shown in \cite{baf}, the energy efficiency (and hence diversity) becomes far more important.

Thus conventional schemes like Amplify and Forward (AF) and Decode and Forward (DF) \cite{laneman} become sub optimal at LSNRs because they are inefficient in the transfer of energy to the network.

In \cite{baf}, a novel Bursty AF (BAF) has been proposed that achieves full diversity at LSNRs. The max-flow min-cut bound was also derived and the performance of BAF was shown to achieve the bound upto a first order approximation.

When the source knows the source-relay channel gain, the performance can be further improved. This method was proposed in \cite{as} where the authors proposed to switch between BAF and DF schemes based on the source-relay channel. Usually the requirement of CSI at transmitter involves a feedback channel which means extra resources and more system complexity. Using the system model of \cite{laneman} however, we see that for any realistic system with two cooperating mobiles, the mobiles must switch their roles as source and relay. Thus the source-relay channel information acquired while the mobile is acting as a relay can be used in the next block while acting as source thereby obviating the need for feedback. 

We also show that this channel information at the source allows us to use the selection relaying protocol. In a selection relaying protocol like Selection Decode and Forward (SDF), the relay transmits only when it is able to decode the received signal completely (i.e.\ when the the source-relay channel exceeds a certain threshold) otherwise the source simply repeats the transmission. Thus SDF also transmits energy continuously into the network thereby achieving full diversity.

\section{System Model} \label{sysmo}
\begin{figure}[ht]
\begin{center}
     \includegraphics[width=2.2in]{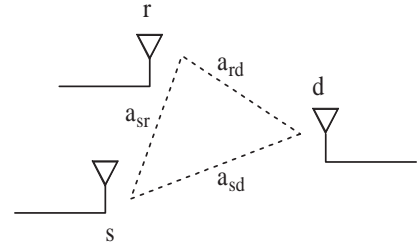} \\
      \caption{System Model} 
      \label{ant}
\end{center}
\end{figure}

\begin{figure}[tbph]
  \begin{center}
    {\includegraphics[width=3.0in]{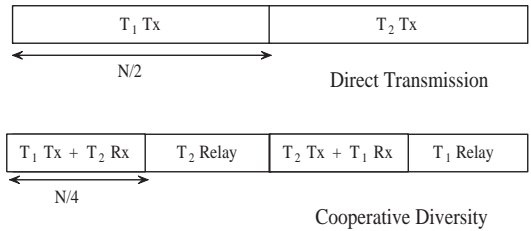}}
      \caption{Channel Allocation for equal data case}
      \label{alloc}
\end{center}
\end{figure}

We consider a system model similar to the one suggested in
\cite{laneman}, consisting of two transmitting terminals and one receive terminal as shown in Fig. \ref{ant}.
The notation, baseband equivalent model and the channel allocation diagram is same as in \cite{laneman}. Thus for direct transmission,
\begin{align}
y_d[n] &= a_{sd}x_s[n] + z_d[n] & 1 \leq n \leq N/2
\end{align}
where N is the total block length (in number of symbols) and
$x_s[n]$ denotes the signal transmitted by the source at time $n$.
The noise term $z_j$ for $j \in \{s,r,d\}$ are also zero mean
circularly symmetric complex Gaussian random variables with power
spectral densities of $N_0$. As in \cite{laneman}, the  subscript is
indicative of the respective terminal (source, relay or
destination). Similarly, for the other terminal, $N/2 + 1 \leq n
\leq N$ as shown in Fig. \ref{alloc}.
For the case of cooperative diversity,
\begin{align} \label{dte}
y_d[n] &= a_{sd}x_s[n] + z_d[n] & 1 \leq n \leq N/4 \\
y_r[n] &= a_{sr}x_s[n] + z_r[n] & 1 \leq n \leq N/4 \\
y_d[n] &= a_{rd}x_r[n] + z_d[n] & N/4+1 \leq n \leq N/2
\end{align}
where $a_{sr}$, $a_{sd}$ and $a_{rd}$ are the channels between
source-relay, source-destination and relay-destination respectively.
Statistically these are modeled as zero mean, independent,
circularly symmetric, complex Gaussian random variables with
variances $\sigma_{ij}$ where $i \in \{s,r\}$ and $j \in \{r,d\}$.
As in \cite{laneman}, the receiver is assumed to have perfect knowledge of the channel gain which remain constant for $N$ symbol intervals (i.e.\ time taken to transmit one block). From now on we will denote $g_1 = \left|a_{sd}\right|^2$, 
$g_2 = \left|a_{rd}\right|^2$ and $h = \left|a_{sr}\right|^2$

Further, similar to \cite{laneman,dmtgamal,baf}, we assume similar implementation
constraints, namely half duplex channel, absence of CSI at
transmitter, and power constraints given by
\begin{align}
\rho &\triangleq \frac{P_s}{N_0}
\end{align}
where $P_s$ is the power of each symbol. Note that similar to \cite{laneman}, CSI about the source-relay channel
is still available to both source and relay as explained earlier.

\section{Selection Decode and Forward in Low SNR}
The Selection Decode and Forward (SDF) scheme was first proposed in \cite{laneman}, where its HSNR behavior was analyzed. Here we evaluate the performance of SDF in LSNR and show that it performs better than any of the existing schemes.

The importance of cooperative diversity at LSNR was first analyzed in \cite{baf}, where the authors showed that the impact of fading and of diversity on the capacity, is much more significant at LSNR while multiplexing gain plays little role. They further showed that the Amplify and Forward (AF) and Decode and Forward (DF) schemes perform poorly. In the AF scheme, the noisy signal received at the relay is amplified and retransmitted to the destination. This scheme fails at LSNR because the relayed signal received at the destination is often too noisy to give diversity advantage. Based on this observation, the authors suggested the Bursty AF (BAF) protocol which by transmitting at low duty cycles and low rates actually overcomes this disadvantage. 

Another interesting scheme that was analyzed was DF. It was shown that although DF gives full diversity, it does not achieve the max-flow min-cut bound. The authors argued that achieving the bound requires continuous transfer of energy in the channel irrespective of the channel gains. The DF scheme, on the other hand, transmits in the second time slot only if it is able to correctly decode the information received from the source in the first slot. The resulting discontinuity in energy transfer inevitably reduces the average energy transfered to the channel, resulting in an increase in outage probability. 

To overcome this problem of discontinuous energy transfer, we propose the use of SDF, where in the event of relay being unable to decode the source signal, the source must retransmit its signal. This makes the transfer of energy continuous and as a result helps achieving better performance. In fact for a higher order analysis, the SDF protocol performs even better than the BAF. 

A similar adaptive scheme (AS) that switches between BAF and DF was proposed recently in \cite{as}. The scheme utilizes source-relay channel gain at the receiver and performs better than the BAF scheme.

\subsection{Performance at Low SNR}
A higher order analysis of various protocols at LSNR was first done in \cite{as}. Taking $\alpha = 2R/\rho$ for consistency with the system model, the results of \cite{as} are repeated here for convenience. Also for the sake of comparison, we assume all channels to have unit variance (i.e.\ $\sigma_{i} = 1$ $\forall$ $i$). 

The outage probability of the BAF and AS protocols are given by
\begin{align}
P_{BAF} &= \alpha^2 - \frac{2}{3}\alpha^3\log \alpha \\
P_{AS} &= \alpha^2 + 2\alpha^3 
\end{align}
where similar to \cite{baf}, $\alpha \rightarrow 0$ as $\rho \rightarrow 0$. The derivation of above expression for BAF is very similar to the one outlined in \cite{baf} except that we consider a higher order term in all our approximations. We now derive the performance of SDF scheme. The mutual information between the source and the destination (see \cite{cover}) was derived in \cite{laneman} and is given by,
\begin{equation} \label{sdf}
I_{SDF} = \begin{cases} \frac{1}{2}\log (1 + 2\rho g_1) & \frac{1}{2}\log(1 + h\rho) <R \\
\frac{1}{2}\log (1 + \rho [g_1+g_2]) & \frac{1}{2}\log(1 + h\rho) >R
\end{cases}
\end{equation}
where the factor of 2 appears because of the repetition coding of the source message. The outage probability is now given by,
\begin{align}
P\{I_{SDF}<R\} &= P\{g_1<\alpha/2\}P\{h<\alpha\} \nonumber\\
&+ P\{g_1+g_2<\alpha\}P\{h>\alpha\} \nonumber\\
&= \frac{\alpha^2}{2} + \frac{\alpha^2}{2} \nonumber\\
&= \alpha^2
\end{align}
A higher order analysis can also be done similarly and gives,
\begin{equation} \label{3o}
P_{SDF} = \alpha^2 - \frac{29}{24}\alpha^3
\end{equation}
We see that the SDF performs better than the BAF protocol. In fact if we were to derive the expression for max-flow min-cut bound upto higher order, it would turn out to be,
\begin{equation}
P_{LB} \geq \alpha^2-\alpha^3
\end{equation}
which is higher than SDF. This is because this max-flow min-cut bound derived in \cite{baf} assumed that the transmitters had no CSI (thus it assumed the independence between the bits transmitted at the source and the destination, see \cite[Appendix III]{baf} for details) which is not the case with the SDF protocol. 

\begin{figure}[tbph]
\begin{center}
      \includegraphics[width=3.5in]{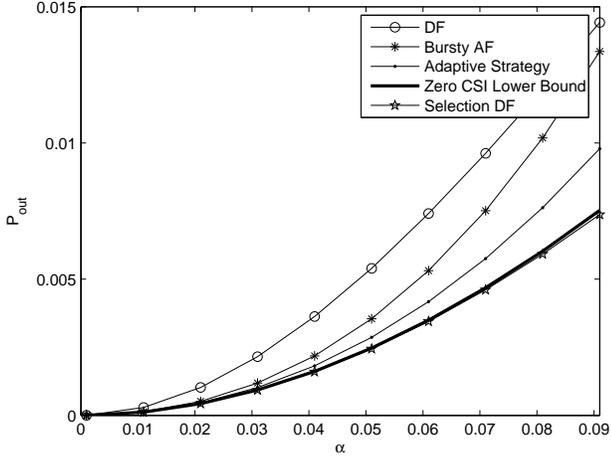} \\
      \caption{Comparison of various schemes at low SNR. $\alpha = 2R/\rho$} 
      \label{comp}
\end{center}
\end{figure}

For comparison we also show the outage probabilities in Fig. \ref{comp} where the improvement provided by SDF over other schemes can easily be seen. 

\section{Energy optimization}
As has already been pointed out in \cite{baf}, energy is a treasured resource at LSNR. Thus energy optimization becomes far more important here. In \cite{baf} an energy optimization was done with respect to the amount of energy allocated to different slots for the max-flow min-cut bound. However as we have seen, the bound has been achieved only approximately. We should therefore optimize the energy allocation with respect to the outage probability itself rather than the bound. 

Assume that a fraction $x$ of the total power is devoted to direct transmission and fraction $1-x$ to relayed (or direct as may be the case) transmission. Thus the SNR at destination is $2x\rho$ in first slot and $2(1-x)\rho$ in the second slot.  The outage probability is now given by,
\begin{align}
P\{I_{SDF}<R\} &= P\{g_1<\alpha/2\}P\{2hx<\alpha\} \nonumber\\
&+ P\{2g_1x+2g_2(1-x)<\alpha\}P\{2hx>\alpha\} \nonumber \\
&= \frac{\alpha^2}{4x} + \frac{\alpha^2}{8x(1-x)} \nonumber \\
&= \frac{\alpha^2}{8} \left(\frac{3-2x}{x(1-x)}\right)
\end{align}
using results from \cite{baf}. Minimizing the above expression in $x$, we get
\begin{equation}
x_{opt} = \frac{3}{2}-\frac{\sqrt{3}}{2} \approx 0.634
\end{equation}
Notice that the corresponding expression for max-flow min-cut bound (given in \cite{baf}) when evaluated for symmetric case gives an optimum value of $x_{opt} \approx 0.667$ which is slightly different from above. Similar difference is expected for non-symmetric case as well, particularly when asymmetry is large (ie.\ for the case when $g_1$, $g_2$ and $h$ differ considerably).

\section{Conclusion and Future work}\label{conc}
The optimality of Selection Decode and Forward (SDF) was analyzed at low signal-to-noise ratio (LSNR) and shown to be better than the recently proposed bursty amplify and forward and the adaptive schemes. Further we showed that energy optimization for SDF yield slightly different results from that of max-flow min-cut bound because of the inherent sub optimality.

\appendix
\label{A}
Here we derive the expression for the outage probability of SDF scheme in terms of $\alpha$ using second order approximations. We start with the general case assuming variances of $a_{sd}$, $a_{rd}$ and $a_{a_{sr}}$ to be $\sigma_{sd}$, $\sigma_{rd}$ and $\sigma_{sr}$ respectively. Using the expression in \eqref{sdf}, the outage probability is given by,
\begin{align} \label{psdf2}
P\{I_{SDF}<R\} &= P\{g_1 < \frac{e^{2R}-1}{2\rho}\}P\{h < \frac{e^{2R}-1}{\rho}\} \nonumber\\
&+ P\{g_1+g_2 < \frac{e^{2R}-1}{\rho}\}P\{h > \frac{e^{2R}-1}{\rho}\}
\end{align}
Now $g_1$, $g_2$ and $h$, as defined before, are exponentially distributed random variables with means $\sigma_{sd}$, $\sigma_{rd}$ and $\sigma_{sr}$ respectively. Therefore $g_1+g_2$ is exponentially distributed with its cumulative distribution function given by,
\begin{equation}
P\{g_1+g_2 < x\} = 1-e^{-x}(1+x)
\end{equation}
Using these results in \eqref{psdf2}, we get
\begin{align} \label{sdfexp}
P\{I_{SDF}<R\} &= \left(1-e^{\frac{1-e^{2R}}{2\rho}}\right)\left(1-e^{\frac{1-e^{2R}}{\rho}}\right) \nonumber\\
&+ \left(1-e^{-\frac{e^{2R}-1}{\rho}}\left(1+\frac{e^{2R}-1}{\rho}\right)\right)\left(e^{\frac{1-e^{2R}}{\rho}}\right)
\end{align}
Since $R$, $\rho$ and $\frac{R}{\rho}$ approach zero, we may simply expand \eqref{sdfexp} using a Taylor series approximation and obtain ,
\begin{equation}
P\{I_{SDF}<R\} = \left(\frac{2R}{\rho}\right)^2 - \frac{29}{24}\left(\frac{2R}{\rho}\right)^3 + O[\left(\frac{2R}{\rho}\right)^4]
\end{equation}
where we have taken $R \rightarrow 0$. Now setting $\alpha = 2R/\rho$ as before, we obtain the desired result \eqref{3o}.


\begin{thebibliography}{99}

\bibitem{baf}
A.\ Salman Avestimehr and David N.\ C.\ Tse,
\newblock {\em Outage Capacity of the Fading Relay Channel in the Low SNR Regime },
\newblock IEEE Trans. Info. Theory, Vol. 51, No. 9, pp 3284-89 Sept. 2005

\bibitem{laneman}
J. Nicholas Laneman, David N. C. Tse and Gregory W. Wornell,
\newblock {\em Cooperative Diversity in Wireless Networks: Efficient Protocols and Outage Behavior },
\newblock IEEE Trans. Info. Theory, vol.50, pp.3062-3080, Dec. 2004.

\bibitem{dmtgamal}
Kambiz Azarian, Hesham El Gamal and Philip Schniter, {\em ``On the Achievable Diversity–Multiplexing Tradeoff in Half-Duplex Cooperative Channels,''} IEEE Trans on Info. Theory, Vol. 51, No. 12, pp 1073-1096, pp. 4152-4172, Dec. 2005.

\bibitem{dmttse}
Lizhong Zheng and David N.\ C.\ Tse, {\em ``Diversity and Multiplexing: A Fundamental Tradeoff in Multiple-Antenna Channels,''} IEEE Trans on Info. Theory, Vol. 49, No. 5, pp 1073-1096, May 2003.

\bibitem{as}
Masoud Sharif, Venkatesh Saligrama, George Atia,
\newblock  Outage Capacity of Relay Channels in Low SNR: An Adaptive Strategy, submitted to CTW 2006.

\bibitem{cover}
Thomas M. Cover, Joy A. Thomas,
\newblock {\em ``Elements of Information Theory },
\newblock John Wiley \& Sons, Inc., 1991.

\end{thebibliography}
\end{document}